\documentclass[twocolumn]{revtex4}
\usepackage{graphicx}
\usepackage{epsfig}
\newcommand{\bra}[1]{\langle {#1} |}
\newcommand{\ket}[1]{| {#1} \rangle}
\newcommand{\ketn}[1]{ {#1} \rangle}
\newfont{\boldit}{cmbxti10}
\begin{document}

\title{Multiscale model of electronic behavior and localization
in stretched dry DNA}

\author{Ryan L.~Barnett$^{1,
\footnote{Present address: 
Department of Physics, California Institute of Technology, 
Pasadena, CA 91125.}
}$, 
Paul Maragakis$^{2,
\footnote{Present address: 
D.E. Shaw Group, 120 West Forty-Fifth St., New York, NY 10036.}
}$, 
Ari Turner$^{1}$, Maria Fyta$^{1}$, and Efthimios Kaxiras$^{1,3}$}

\affiliation{ $^{1}$Department of Physics, Harvard University,
Cambridge, MA 02138\\
$^{2}$Department of Chemistry and Chemical Biology, Harvard
University, Cambridge, MA 02138\\
$^{3}$School of Engineering and Applied Sciences, Harvard
University, Cambridge, MA 02138
 }

\begin{abstract}
When the DNA double helix is subjected to external forces it can
stretch elastically to elongations reaching 100\% of its natural
length. These distortions, imposed at the mesoscopic or
macroscopic scales, have a dramatic effect on
electronic properties at the atomic scale and on electrical
transport along DNA. Accordingly, a multiscale approach is
necessary to capture the electronic behavior of the stretched DNA helix.
To construct such a model, we begin with
accurate density-functional-theory calculations for electronic
states in DNA bases and base pairs in various relative
configurations encountered in the equilibrium and stretched forms.
These results are complemented by semi-empirical quantum
mechanical calculations for the states of a small size [18
base pair poly(CG)-poly(CG)] dry, neutral DNA sequence, using previously
published models for stretched DNA. The
calculated electronic states are then used to parametrize an
effective tight-binding model that can describe electron hopping
in the presence of environmental effects, such as the presence of
stray water molecules on the backbone or structural features of
the substrate. These effects introduce disorder in the
model hamiltonian which leads to electron localization. The 
localization length is smaller by several orders of magnitude in
stretched DNA relative to that in the unstretched structure.
\end{abstract}
\maketitle

\section{Introduction}

Soon after Watson and Crick's discovery of the DNA double-helix
structure \cite{WATSON1953}, Eley and Spivey \cite{ELEY1961}
introduced the notion of efficient charge transport along the
stacked $\pi$ orbitals of the bases. The mechanism of charge
transport has been the subject of numerous studies in the
intervening years, with renewed interest fuelled recently by both
biological and technological considerations. Over a decade ago,
Barton and co-workers observed distance-independent charge
transfer between DNA-intercalated transition-metal complexes
\cite{MURPHY1993} and argued that it would be relevant for biology
and biotechnology. More recent electron transport experiments on
DNA have yielded widely varying results, showing alternatively
insulating behavior
\cite{BRAUN1998,DEPABLO2000,STORM2001,ZHANG2002,BABA2006},
semiconducting behavior \cite{PORATH2000}, Ohmic conductivity
\cite{FINK1999,CAI2000,TRAN2000,COHEN2005}, and proximity induced
superconductivity \cite{KASUMOV2001}. The large number of relevant
variables endemic to such experiments, like the
DNA-electrode contact, and the rich variety of
structures that DNA can assume, are the causes of variability in
the experimental measurements 
(for a recent review of transport theory and experiments see
Ref.~\cite{ENDRES2004}).

Specifically, there is a large diversity of the DNA forms
in terms of its composition, length, and structure.
Experiments done long ago,
suggested that DNA substantially longer than its natural length 
(also referred to as ``overstretched DNA'') can undergo 
a transition to an elongated structure
up to twice the length of relaxed DNA \cite{WILKINS51}. 
This was also confirmed by recent single molecule stretching experiments
\cite{SMITH1996,CLUZEL1996,STRICK2000}, which showed that
the molecule can be reversibly stretched up to 90\% of its natural length.
Such important deformations of the double helix may occur 
in biological environments. Stretching of DNA is also
related to cellular processes, such as transcription and
replication. For example, proteins often induce important
local distortions in the double helix while they diffuse along the
molecule in search of their target sequences.
The electronic and transport properties of DNA are directly
influenced by its different conformations as well as by
environmental factors, such as counterions, impurities or
temperature. A full account of these effects
based on a realistic, atomic scale description of the structure
and the electronic properties challenges the capabilities of
theoretical models.

Theoretical efforts to understand the electronic behavior and
transport in DNA can be divided into two general categories:\\
(i) Model calculations that use effective hamiltonians and master
equations to describe the dynamics of electrons and holes in DNA
(see, for instance, Refs.~\cite{YAMADA2004,IGUCHI2004,Yi2003,Chang2004}).
Recent results \cite{CAETANO2005} have led to considerable insights 
concerning the sequence-independent delocalization of electronic 
states in DNA. The main limitation of such approaches lies in the 
difficulty of determining accurate values for the parameters in
the effective hamiltonians.\\
(ii) {\it Ab initio} calculations that can provide an accurate and
detailed description of the electronic features
\cite{RNBARNETT2001,ARTACHO2003,ALEXANDRE2003}. 
These approaches are typically limited to a small number of atoms due to
computational costs, and cannot readily handle the full complexity
of DNA molecules in various conformations. In particular,
stretching of DNA can induce a very significant deviation from the
B form which is stable under normal conditions in aqueous
solution. Such structural distortions are bound to have a profound
effect on the electronic behavior.  A realistic description of
these effects makes it necessary to handle both the atomic scale
features and the overall state of the macromolecule.

In the present work, we address the problem of DNA stretching
effects on the electronic states and the
electron localization by providing a bridge between the
two extremes of the length scale; 
a similar methodology was recently used to study hole
transfer in DNA \cite{SENTHI05}.
Theoretically, there are different ways of pulling the opposite ends
of the DNA strands, leading to different stretched DNA
forms, which are determined largely by base pair reorientations. 
Here, we use the poly(CG)-poly(CG) structures obtained in the
pioneering study of Lebrun and Lavery \cite{LEBRUN1996}
as the representative structure for stretching effects. This
study modeled the adiabatic elongation of selected DNA molecules
in two modes of stretching, corresponding to pulling on opposite
3'-3' ends or 5'-5' ends of the molecule: In the 3'-3' stretching
mode, the DNA helix is unwound leading to a ribbon-like structure, 
while in the 5'-5' stretching mode the DNA helix contracts.

\begin{figure}
\includegraphics[width=3.0in]{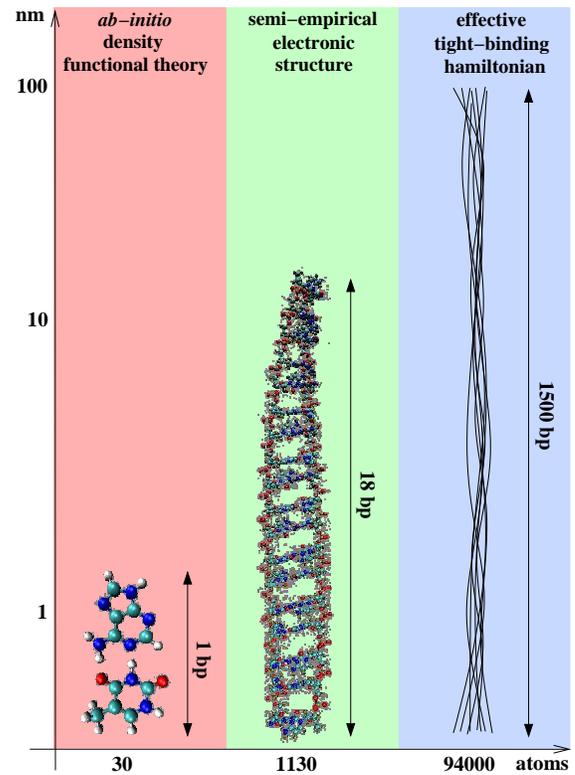}
\caption{Schematic illustration of the different scales included in 
the current multiscale model: The two pictures on the left are
atomistic systems simulated with different computational
approaches ({\it ab initio} density 
functional theory and semi-empirical electronic
structure, resprectively). The picture on the right 
represents a rope composed of DNA molecules, as in experiments
\cite{Heim2004}, which is treated by an effective tight-binding 
hamiltonian constructed from the atomistic scale calculations. } 
\label{Fig:multiscDNA}
\end{figure}

We begin with a set of detailed
calculations for the electronic structure of DNA bases (A,T,C,G)
and representative base pairs (AT-AT, CG-CG, AT-CG, CG-GC)
in various relative configurations, as they are likely to
appear in the stretched forms,
These calculations are based on
density-functional theory \cite{DFT,LDA} and serve to set the
stage for more extensive calculations which employ successive
levels of approximations necessary to handle the computational
demands. 
Specifically, we extract the salient features of electronic 
structure of the individual DNA bases and base pairs from the {\it ab initio}
calculations; these are compared to an efficient and realistic semi-empirical
model \cite{ELSTNER1998a}, in order to establish the validity of 
the latter approach. At this intermediate scale, we consider an 
18 base pair poly(CG)-poly(CG) DNA sequence which has been stretched 
by 30\%, 60\% and 90\% relative to the natural length of the 
unstretched B form. The atomic structure of these forms has been 
established by Lebrun and Lavery \cite{LEBRUN1996}, using empirical 
interatomic potentials. 
We next use the information from this
approximate description to build an effective hamiltonian for the
electronic behavior at much larger scales. This allows us to
describe electron localization, due to the combined effects of 
stretching and environmental factors, over mesoscopic to macroscopic 
length scales. The essence of the approach
and the different scales involved are shown schematically
in Fig. \ref{Fig:multiscDNA}.
We emphasize that we address here issues related only to dry
and neutral DNA structures, where the negatively charged groups
on the backbone are passivated by protons, 
conditions that are relevant to the experiments 
we consider for comparison to our theoretical results;
water molecules
or counterions (such as Na$^+$) are not considered in our 
calculations. 

\section{Theoretical Methods}

\subsection{{\boldit Ab initio} calculations}

As our first step toward establishing the electronic behavior of
dry, neutral DNA, we study the nature of electronic states in individual 
bases and in base pairs. For these calculations we used three different
implementations of density-functional theory \cite{DFT}: a method
that uses atomic-like orbitals as the basis \cite{SIESTA}, one
that uses plane waves \cite{VASP} and a third that uses a
real-space grid \cite{HARES}. In all three approaches, we used the
same exchange-correlation functional in the local-density
approximation \cite{LDA}, for consistency and simplicity. More
elaborate approximations to exchange-correlation effects, such as
the generalized gradient approximation \cite{GGA}, do not provide
any improvement in describing the physics of these weakly
interacting units.
In each method we used pseudopotentials to
represent the atomic cores, of the Trouiller-Martins
type \cite{Trouiller} in SIESTA, the Vanderbilt ultrasoft
type \cite{Vanderbilt} in VASP and the Hammann-Schluter-Chiang type
\cite{HSC} in HARES, with computational parameters (number of
orbitals in basis, plane-wave kinetic energy cutoff and grid
spacing) that ensure a high level of convergence. These
calculations provide a thorough check on the consistency of
various computational schemes to reproduce the electronic features
of interest. The results are in excellent agreement across the
three approaches. Since in these calculations there are no
adjustable parameters, we refer to them in the following as {\it
ab initio} results.

\subsection{Construction of semi-empirical  model}

The stretched forms contain a large number of atoms, typically 
beyond what can be efficiently treated with the {\it ab initio}
methods used for the DNA bases and base pairs.
Accordingly, for the electronic structure calculations of these
structures we use an efficient semi-empirical quantum-mechanical
approach which employs a minimal basis set \cite{ELSTNER1998a}.
The consistency of this approach is then verified against 
the {\it ab-initio} calculations. Within the semi-empirical scheme, 
the electronic eigenfunctions are expressed as
\begin{equation}
\label{wf}
\ket{\psi^{(n)}}=\sum_{\nu}c^{(n)}_{\nu}\ket{\varphi_{\nu}}
\end{equation}
where the basis set $\ket{\varphi_{n}}$ includes the $s$ and $p$ atomic 
orbitals for each atom in the system. 
The coefficients $c^{(n)}_{\nu}$ are numerical constants,
with $|c^{(n)}_{\nu}|^2$ giving the weight of orbital
$\ket{\varphi_{n}}$ to the electronic wavefunction.
This method uses a second
order expansion in the electronic density to obtain the total
energy and takes into account self-consistently charge transfer
effects which are important for biological systems. The method
gives results for the band gaps that are in excellent agreement
with those of the {\it ab initio} approaches described above
(see Refs.~\cite{MARAGAKIS2002,DEPABLO2000}).

The highest 
 occupied and lowest unoccupied molecular orbitals (HOMO and LUMO, 
 respectively, also referred to collectively as ``frontier
 states'' in the following) are extended over the entire structure 
in Bloch-like wave functions. In order to describe electron hopping 
and localization, we need to express these in terms of a basis of 
Wannier-like states that are localized on the individual bases.
To this end, we construct maximally localized states on single 
base pairs by taking linear combinations of the HOMO and LUMO 
states from the wavefunctions of Eq.~(\ref{wf}). The maximally 
localized states will then be used to calculate the hopping 
parameters in the effective 1D hamiltonian. Using the extended 
electronic states $\ket{\psi^{(n)}}$ of the frontier states, with 
corresponding energies $\varepsilon_{(n)}$, we define the maximally 
localized states $\ket{\tilde{\psi}^{(i)}}$ through the unitary
transformation
\begin{equation}
\label{Eq:transformation} \ket{\tilde{\psi}^{(i)}}=\sum_{n}
\bra{\psi^{(n)}}\ketn{\tilde{\psi}^{(i)}} \ket{\psi^{(n)}}
\end{equation}
which minimizes the sum of the variances 
\begin{equation}
\label{Eq:variance} \zeta= \sum_{i}\left(\bra{\tilde{\psi}^{(i)}}
\hat{z}^2
 \ket{\tilde{\psi}^{(i)}}-\bra{\tilde{\psi}^{(i)}}
{\hat{z}}\ket{\tilde{\psi}^{(i)}}^{2} \right)
\end{equation}
under the constraint
$\bra{\tilde{\psi}^{(i)}}\tilde{\psi}^{(j)}\rangle=\delta_{ij}$
where $z$ is the position along the helical axis.  
Similar and more general methodologies have been developed
in the past for obtaining maximally localized states from extended 
ones \cite{MARZARI97,SGIAVORELLO01}. Due to the
invariance of the trace, the first term in Eq.~(\ref{Eq:variance})
is independent of the unitary transformation and the problem is
simplified to one of maximizing the second term on the right-hand
side with the same orthonormality constraint. Carrying out the
minimization, we arrive at the equation
\begin{equation}
\label{Eq:iteration} \bra{\tilde{\psi}^{(n)}}\hat{z}
\ket{\tilde{\psi}^{(m)}}(z_{n}-z_{m})=0
\end{equation}
where
\begin{equation}
z_{n}=\bra{\tilde{\psi}^{(n)}} \hat{z} \ket{\tilde{\psi}^{(n)}}.
\end{equation}
By inspection, we see that $\zeta$ is maximized when $z_{n}=z_{m}$
for all $m$ and $n$, corresponding to maximally delocalized
states. On the other hand, $\zeta$ is minimized when the states
$\ket{\tilde{\psi}^{(n)}}$ are the eigenfunctions of the position
operator $\hat{z}$ within the HOMO or LUMO subspace. Therefore,
the problem is further reduced to constructing and diagonalizing
the matrix
\begin{equation}
M_{nm} = \bra{\psi^{(n)}}\hat{z} \ket{\psi^{(m)}}
\end{equation}
which has the eigenvectors
$\bra{\psi^{(n)}}\ketn{\tilde{\psi}^{(i)}}$ that provide the
desired transformation given in Eq.~(\ref{Eq:transformation}). The
eigenvalues $z_{n}$ are the positions of the localized states. To
evaluate the matrix elements  we use the approximation
\begin{eqnarray}
\bra{\psi^{(n)}}\hat{z}\ket{\psi^{(m)}} & = &
\sum_{\mu\nu}c^{(n)*}_{\mu}c^{(m)}_{\nu}
\bra{\varphi_{\mu}}\hat{z}\ket{\varphi_{\nu}}
\nonumber \\ & \approx &
\sum_{\mu\nu}c^{(n)*}_{\mu}c^{(m)}_{\nu}S_{\mu\nu}z_{\mu\nu}
\end{eqnarray}
where $S_{\mu\nu}=\bra{\varphi_{\mu}}\varphi_{\nu}\rangle$ is the
overlap matrix between the two atomic orbitals and 
 $z_{\mu\nu}=\frac{z_\mu + z_\nu}{2}$ is the average $z$-value 
for the atoms located at sites given by the labels $\mu$ and $\nu$. 
Once the localized states are constructed, the hopping parameters 
can be computed as
\begin{equation}
t_{ij}= \bra{\tilde{\psi}^{(i)}}{\cal H}\ket{\tilde{\psi}^{(j)}}=
\sum_{n} \varepsilon^{(n)}
\bra{\tilde{\psi}^{(i)}}\ketn{\psi^{(n)}}
\bra{\psi^{(n)}}\ketn{\tilde{\psi}^{(j)}}
\end{equation}
recalling that the quantities
$\bra{\psi^{(n)}}\ketn{\tilde{\psi}^{(i)}}$ are determined from
the transformation described above.

\begin{figure}
\includegraphics[width=2.0in]{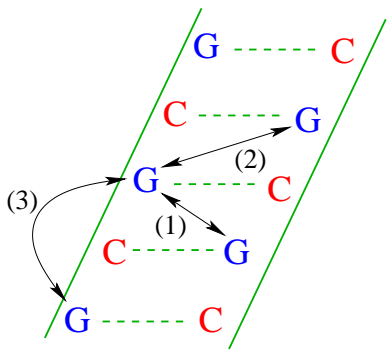}
\caption{Schematic depiction of electron hopping in
poly(CG)-poly(CG) DNA for the HOMO state. The hopping matrix
elements $t_i$ are denoted by the indices $(i) = (1),(2),(3)$.
Electrons are localized on the G bases. For the LUMO state, the
hopping is similar with electrons localized on the C bases.}
\label{Fig:schem}
\end{figure}

Having defined the maximally localized states in terms of the
electronic wavefunctions from the all-atom calculations, we next
produce an effective tight-binding hamiltonian, which allows us
to study electron hopping along the DNA double helix. 
This approach has also been used in a recent study 
on functionalized carbon nanotubes \cite{LEE05}.
In our
effective hamiltonian, we consider hopping between first and second
neighbors along the helix, and denote the hopping matrix
elements according to the scheme shown in Fig.~\ref{Fig:schem} for
the HOMO state of the poly(CG)-poly(CG) structure (all other
frontier states involve exactly the same type of hopping matrix
elements):
\begin{eqnarray}
\label{Eq:Hopping}
{\cal H}= \varepsilon \sum_{n}
c_{n}^{\dagger}c_{n}&+& t_{1}\sum_{n \:  {\rm even}}
\left(c_{n}^{\dagger}c_{n+1}+c_{n+1}^{\dagger}c_{n} \right)
\nonumber \\
&+& t_{2}\sum_{n \:  {\rm odd}}
\left(c_{n}^{\dagger}c_{n+1}+c_{n+1}^{\dagger}c_{n} \right)
\nonumber \\
 &+& t_{3}\sum_{n } \left(c_{n}^{\dagger}c_{n+2}
+c_{n+2}^{\dagger}c_{n} \right)
\end{eqnarray}
where $n$ represents the $n^{th}$ base pair along the helical axis
and we have neglected spin indices because they are unimportant
for our analysis. 
Note that there is a difference between hopping elements connecting
even and odd sites to their neighbors ($t_1$ and $t_2$ terms
in the effective hamiltonian of Eq.~(\ref{Eq:Hopping})), due 
to the asymmetry in the structure illustrated in Fig.~\ref{Fig:schem}.
Performing a Fourier transform on the electron creation and 
annihilation operators
\begin{equation}
c_{k}=\frac{1}{\sqrt{N}}\sum_{n} e^{-i k n} c_{n}
\end{equation}
gives a hamiltonian which has coupling between momenta $k$ and
$k+\pi/a$. By doubling the unit cell (and reducing the Brillouin
Zone by a factor of two), this can finally be diagonalized to
obtain the eigenvalues
\begin{eqnarray}
E_{k}^{\pm}=\varepsilon+2 t_{3} \cos(2k) \pm&
\sqrt{t_{1}^{\;2}+t_{2}^{\;2}+2 \:t_{1}t_{2}\cos(2k)}
\end{eqnarray}
with the momentum sum carried out over the reduced Brillouin Zone.
With these expressions for the band structure energies, the
density of states (DOS)
\begin{equation}
g(\omega)=\frac{1}{N}\sum_{k,n} \delta(\omega-E_{k}^{(n)})
\label{DOS}
\end{equation}
can be readily obtained. These quantities are essential in
describing electron localization along the DNA double helix under
different conditions.

\subsection{Disorder and Localization length}
\label{Sec:Disorder}

In order to quantify
the amount of localization that is expected in stretched DNA
forms, we add a term to the hamiltonian in Eq.~(\ref{Eq:Hopping}) of
the form
\begin{equation}
{\cal H}_{\rm dis}= \sum_{n} U_{n} c_{n}^{\dagger} c_{n}
\end{equation}
which is meant to emulate disorder arising from a variety of
sources such as interaction of the DNA bases with stray water
molecules and ions, or interaction with the substrate. $U_{n}$ are
uncorrelated random energy variations chosen according to a
Gaussian distribution of zero mean and width $\gamma$
\begin{equation}
P(U)=\frac{1}{\gamma \sqrt{2 \pi}} \exp\left(-\frac{U^2}{2\gamma^2}\right).
\label{PU_disorder}
\end{equation}
Once the disorder hamiltonian is constructed with a specific set
of random on-site energies, by direct diagonalization we find the
eigenstates $\ket{\Psi^{(i)}}$ of ${\cal H}+{\cal H}_{\rm dis}$
(we use capital symbols to denote the new wavefunctions 
from the hamiltonian that includes the disorder term)
and then calculate the localization length defined as
\begin{equation}
L_{i}=\left[\bra{\Psi^{(i)}}\hat{n}^2 \ket{\Psi^{(i)}}-
\bra{\Psi^{(i)}}\hat{n}\ket{\Psi^{(i)}}^{2}\right]^{1/2}
\label{loc_length}
\end{equation}
where  
\begin{equation}
\hat{n}=\sum_{n}n c_{n}^{\dagger} c_{n}.
\end{equation}
For a single-hopping model with weak disorder, the localization
length scales as $L \sim \left(t/\gamma\right)^2$ for electrons
near the middle of the band \cite{THOULESS1972},
with $t$ the hopping matrix element which determines
the band width.  The more
complicated effective hamiltonian considered here is not amenable
to simple analytic treatment.

\section{Results and Discussion}

We begin our discussion with an overview of electronic states in
single bases and isolated base pairs. The structure of the base 
pairs is shown in Fig. \ref{Fig:structures} with the atoms in 
each base labeled for future reference. These calculations will 
set the stage for a proper interpretation of the behavior in the  
stretched and unstretched dry, neutral DNA helix.

%\subsection{Individual bases and base-pairs}
\subsection{Frontier states}

\begin{figure}
\includegraphics[width=4in]{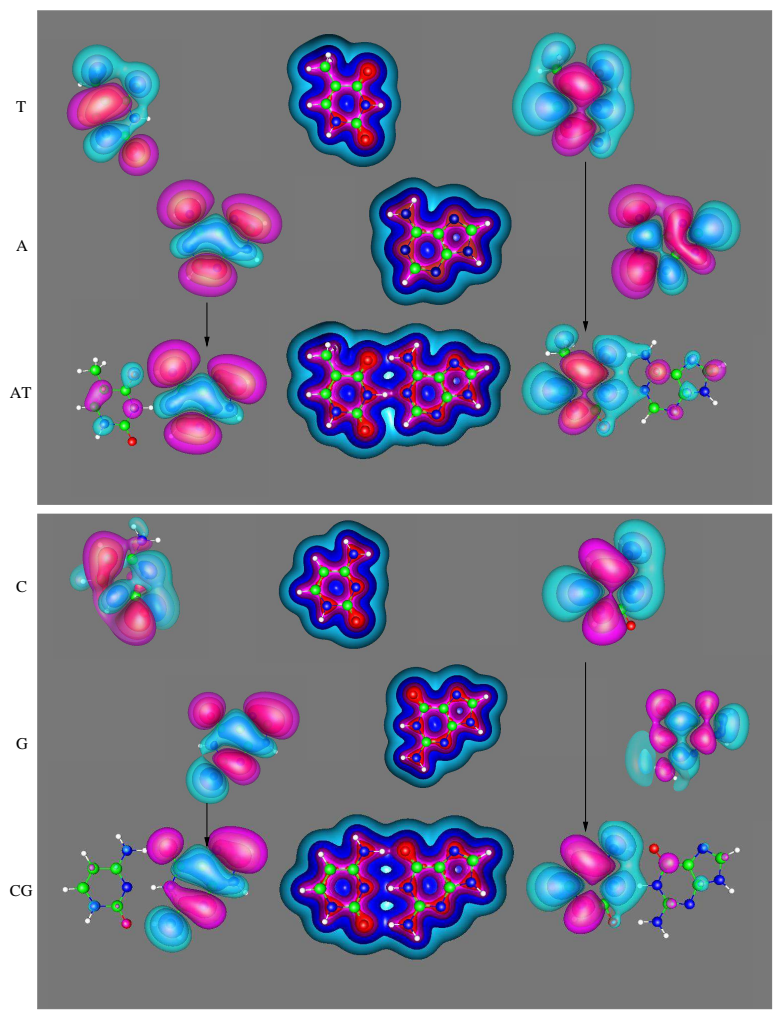}
\caption{The frontier states in the base pairs and their
identification with corresponding orbitals in the isolated bases.
The middle figure in each panel shows the total charge density 
on the plane of the base pair, with higher values of the charge density 
in red and lower values in blue. The figure on the left shows the 
HOMO state and the figure on the right shows the LUMO state, where 
red and blue isosurfaces correspond to positive and negative 
values of the wavefunctions. The labels on the left
denote the type of bases and base pairs. 
}
\label{Fig:basepair_states}
\end{figure}

The frontier states in the base pairs are related to only one 
component of the pair for both AT and CG. This is shown in Fig.
\ref{Fig:basepair_states}. Specifically, the HOMO state of the AT
pair is exactly the same as that of the HOMO state of the isolated
A, and the LUMO state of AT the same as that of the isolated T.
Similarly, the HOMO state of CG is identified with that of the
isolated G and the LUMO state with that of the isolated C. Thus, 
the purines (A or G) give rise to the HOMO state, while the pyrimidines 
(T or C) are responsible for the LUMO states of each pair. It is clear 
from the same figure, that essentially all atomic $p_z$ orbitals which 
belong to a purine or pyrimidine contribute to the respective HOMO 
or LUMO $\pi$ state of the base pair. This is in agreement 
with calculations on the optical absorption spectra of DNA bases and 
base pairs \cite{VARSANO2006}. A closer inspection of 
Fig. \ref{Fig:basepair_states} shows that the molecular frontier 
states of both AT and CG can be identified as similar contributions 
(up to sign changes) from specific groups of carbon and nitrogen atoms. 
Specifically, in the purines (A and G) three distinct groups of 
atoms are mainly involved in forming the HOMO orbital and include 
atoms (C8-N7), (C2-N3) and (N1-C6-C5-C4-N9), respectively. In the 
pyrimidines (T and C) the main groups involved in forming the 
LUMO orbital are two, (C4-C5-N1) and (N3-N7-C6). In both base pairs
the atoms that are less involved in the frontier molecular states  
are the carbon atoms that form a double bond with an oxygen atom, 
such as C2 of A and C and the four-fold bonded C7 atom of A.

\begin{figure}
\includegraphics[width=3.5in]{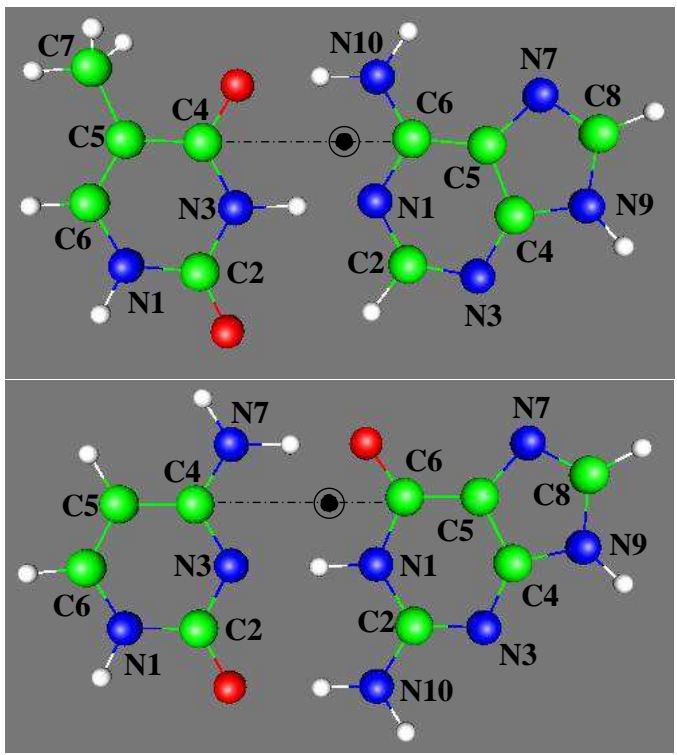}
\caption{The DNA base pairs AT (top) and CG (bottom), 
with the atoms labeled. The purines (A, G) are on the right,
the pyrimidines (T, C) on the left. Atom labeling 
follows standard notation convention \cite{DNAbook}. All 
rotations were performed with respect to the helical axis 
denoted by the black circle (see text).} 
\label{Fig:structures}
\end{figure}

\begin{figure}
\includegraphics[width=3.5in]{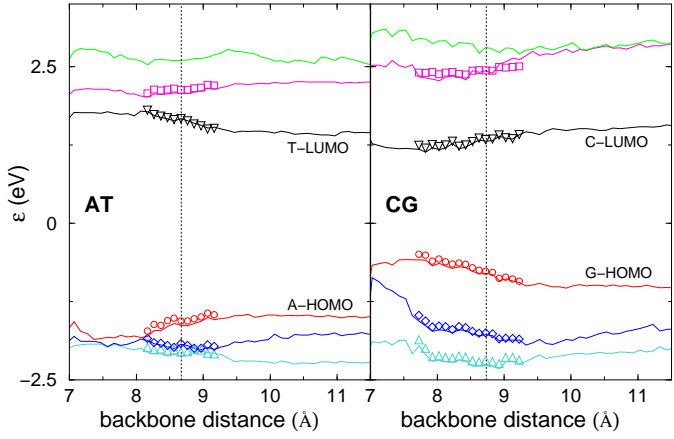}
\caption{Eigenvalues of states in the AT and CG base pairs as a
function of backbone distance. In each case three states are
included above and below the band gap. Lines are results from 
SIESTA calculations, points are results from HARES
calculations (see text). The frontier orbitals in both pairs 
are related to one component of the pair as indicated by the labels.
The equilibrium backbone distance is denoted by a vertical 
dashed line.} 
\label{Fig:backbone_sep}
\end{figure}

The frontier states are very little affected when the two components 
of the base-pair are separated along the direction in
which they are hydrogen-bonded. To demonstrate this, we show in
Fig. \ref{Fig:backbone_sep} the change in the eigenvalues of the
frontier states in AT and CG as a function of the distance between
the two atoms that are bonded to the two backbones (we call this
the backbone distance). For both base pairs the nitrogen atoms 
labeled N1 and N9, are the ones attached to the backbone (see 
Fig. \ref{Fig:structures}). In order to obtain realistic structures,
for each value of the backbone distance we hold the atoms of each
base that are bonded to the backbone fixed and allow all other
atoms to relax fully. These calculations were performed with the
SIESTA code \cite{SIESTA} and the relaxed configurations were used
as input to calculate the electronic structure with the other two
methodologies \cite{VASP,HARES}. In Fig. \ref{Fig:backbone_sep}
we show complete results from the SIESTA calculations and selected
results from one of the other two approaches.

The results of Fig. \ref{Fig:backbone_sep} show clearly that only
in the region where the backbone distance becomes significantly
smaller than the equilibrium value, interaction between the two
bases shifts the eigenvalues of the electronic states appreciably,
but even then the shifts are relatively small for the frontier
states. It is also noteworthy that the band gap of the AT pair is
significantly larger ($\sim 3$ eV) than that of the CG pair ($\sim
2$ eV) and that the frontier states of CG lie within the band gap
of the AT pair. This observation is important because it
indicates that in an arbitrary sequence of base pairs, the
frontier states will be associated with those of the CG pairs.
This statement is verified by calculations of electronic states in
the AT-AT, CG-CG and AT-CG base pair combinations, to which we
turn next.  

For more detailed comparisons, we collect in Table
\ref{Table:eigenvalues} the eigenvalues of the frontier states
for the DNA pairs and the pair combinations, at different 
equilibrium configurations in the three relevant variables, 
the backbone distance, the axial distance and the rotation 
angle. Some results on the CG-GC base pair combination are 
also shown, to allow for comparison to the poly(C)-poly(G) sequence.

\begin{table}
\begin{tabular}{|c|r|r|r|r|}\hline
&min &HOMO & LUMO & gap \\ \hline \hline
 Backbone distance & & & & \\ \hline 
AT&$8.67$~\AA &$ -1.63$  &1.60 & 3.23 \\ 
CG& $8.73$~\AA&$ -0.80$  &1.31 & 2.11 \\ \hline \hline
Axial distance  & & & & \\ \hline 
AT-AT&$3.67$~\AA &$-1.33$ &1.37 & 2.70 \\ 
CG-CG&$3.52$~\AA &$-0.46$ &0.95 & 1.41\\ 
%CG-GC&$3.54$~\AA &$-1.09$ & 1.55 & 2.64 \\ 
AT-CG&$3.36$~\AA &$-0.71$ &1.00 & 1.71 \\ \hline \hline
Rotation angle & & & & \\ \hline
&$36^{o}$ &$-1.48$ &1.58 & 3.06 \\ 
AT-AT&$108^{o}$ &$-1.45$  &1.68 & 3.13 \\ 
&$180^{o}$ &$-1.55$  &1.63 & 3.18 \\ \hline
&$36^{o}$ & $-0.52$  &1.22 & 1.74\\ 
CG-CG&$108^{o}$ &$-0.64$  &1.54 & 2.18 \\ 
&$180^{o}$ &$ -0.94$ &1.60 & 2.54 \\ \hline
&$36^{o}$ & $-0.86$  &1.51 & 2.37 \\ 
CG-GC&$108^{o}$ &$-0.66$  &1.43 & 2.09 \\ 
&$180^{o}$ &$-0.60$ & 1.12 & 1.72 \\ \hline
&$36^{o}$ & $-0.73$  &1.38 & 2.11 \\ 
AT-CG&$108^{o}$ &$-0.59$ &1.27 & 1.86  \\ 
&$180^{o}$ &$ -0.81$  &1.25 & 2.06 \\ \hline 
\end{tabular}
\caption{Eigenvalues (in eV) of the frontier states for the DNA 
base pairs and the base-pair combinations, at the equilibrium 
configurations for the backbone distance,
the axial distance (at zero relative angle of rotation)
and the angle of rotation (at the equilibrium axial distance).
The column labeled ``min'' gives the values of the 
distances and the angle at the equilibrium configurations.
Due to symmetry the values for the minima at rotation angles
larger than 180$^o$
are similar to those given here and are not shown.}
\label{Table:eigenvalues}
\end{table}

\begin{figure}
\includegraphics[width=4in]{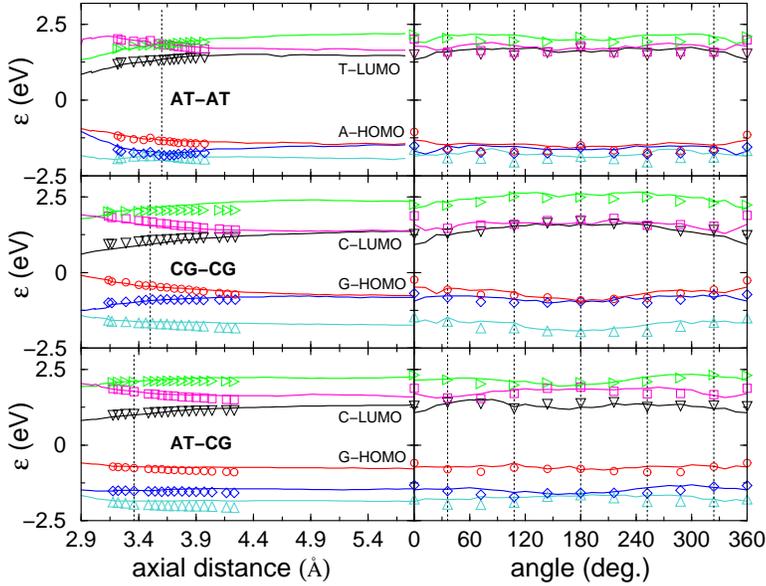}
\caption{Eigenvalues of states in the AT-AT, CG-CG and AT-CG base
pair combinations as a function of the distance along the helical
axis (at zero angle of rotation) and the rotation angle around
the helical axis (at the equilibrium axial distance). Lines are
results from SIESTA calculations, points are results from  
VASP calculations (see text). In each case three states are
included above and below the band gap. The value of the distance
or the rotation angle that correspond to equilibrium
configurations are indicated by vertical dashed lines (there are
five almost equivalent local minima in rotation). As in
Fig. \ref{Fig:backbone_sep}, frontier orbitals are identified 
as the corresponding orbital of one base only.}
\label{Fig:basepair_stack}
\end{figure}

When two base pairs are stacked on top of each other, there are
two degrees of freedom for motion of one relative to the other: a
separation along the helical axis, which we will call axial
distance, and a relative rotation around the helical axis. We
take the helical axis to be that which corresponds to stacking of
successive base pairs in the B form of the DNA double helix. 
According to the notation of Fig.~\ref{Fig:structures}, the 
helical axis for both base-pairs is normal to the line connecting 
atoms C4 and C6 and is closer (about one third of their distance) 
to the purine atom C6.
For each configuration we fix the atoms that are bonded to the
backbone at a given relative position and allow all other atoms to
relax, as was done in the calculations involving the backbone
distance discussed above. In Fig. \ref{Fig:basepair_stack} we show
the behavior of electronic eigenvalues as a function of the axial
distance and the rotation angle. As above, the eigenvalues show
little dependence on these two variables, except for rather small
values of the axial distance which correspond to unphysically
small separation between the two base pairs.

What is also remarkable in the above results, is that in the AT-CG
combination, the frontier states are clearly identified with 
those corresponding to the CG pair exclusively, which has the
smaller band gap (see Fig. \ref{Fig:basepair_stack}). 
Moreover, we note that the band gap of the poly(C)-poly(G) 
sequence, as calculated by the semi-empirical method 
based on a minimal atomic orbital basis \cite{ELSTNER1998a}
is in excellent agreement with the value obtained from the SIESTA
calculation (2.0 eV and 2.1 eV, respectively).
The band gap is expected to be significantly smaller in the case of wet DNA
and in the presence of counterions, 
as shown in Ref.~\cite{GERVASIO2002}, for a Z-DNA helix.
The band gaps between all three {\it ab initio} methods are 
identical within the accuracy of these methods. The nature of 
electronic wavefunctions obtained by the different methods is 
also in good qualitative agreement. Accordingly, in the rest of 
this paper we focus our attention to electron localization in 
the dry, neutral poly(CG)-poly(CG) sequence, and employ the results of the
semi-empirical electronic structure method.

\subsection{Hopping electrons}

\begin{table}
\label{table}
\begin{tabular}{|c|r|r|}\hline 
& HOMO & LUMO \\ \hline
$\varepsilon\;$(eV) & 3.12 & $-0.09$ \\
$t_{1}\;$(meV)& 14.0 & $-0.29$\\
$t_{2}\;$(meV)& 2.60 & 0.04\\
$t_{3}\;$(meV)& 0.09 & 0.26\\ \hline 
\end{tabular}
\caption{Parameters for the on-site ($\varepsilon$) and hopping
matrix elements ($t_i, i=1,2,3$), for the HOMO and LUMO 
states of unstretched poly(CG)-poly(CG) DNA.}
\label{Table:hoppings}
\end{table}

\begin{figure}
\includegraphics[width=4.0in]{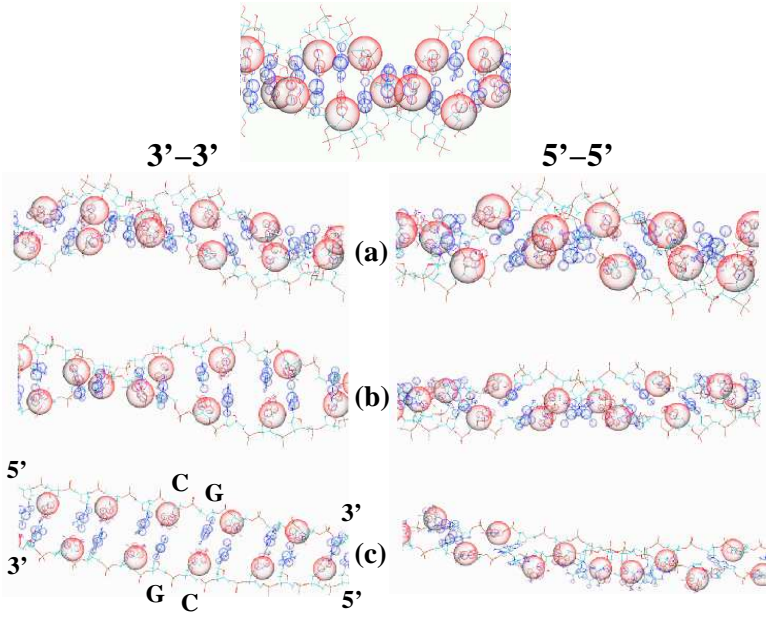}
\caption{The DNA structures for the unstretched (top) and the 
different amounts of stretching in the 3'-3' and the 5'-5' modes 
with features of the frontier orbitals described by the 
blue (HOMO) and red (LUMO) spheres (see text for details). For 
both modes the amount of stretching is (a) 30\%, (b) 60\%, and 
(c) 90\% relative to the unstretched structure, which is 
the B-DNA form.
The 3'-5' orientations of the poly(CG)-poly(CG) sequence
are shown in the left panel at 90\% stretching, where these 
the structure is easier to visualize.
}
\label{Fig:balls}
\end{figure}

In Fig. \ref{Fig:balls}, we show the unstretched and the three 
stretched forms of the poly(CG)-poly(CG) sequences at 30\%, 
60\%, 90\% elongation, along with the features of the 
frontier states. For visualization purposes, we
represent the calculated wavefunction magnitude of the frontier
states by blue (HOMO) and red (LUMO) spheres, centered at the
sites where the atomic orbitals are located. The radius of the
sphere centered on a particular atom is proportional to the
magnitude of the dominant coefficient $|c^{(n)}_{\nu}|^2$ 
at this site (see Eq.~(\ref{wf})), which is essentially
proportional to the local electronic density. It is evident from 
this figure that
the nature of the orbitals themselves,
represented by the radii of the colored spheres,
does not change much in the
different stretched DNA forms, but the {\em overlap} between
orbitals at neighboring bases is affected greatly by the amount of
stretching. For the poly(CG)-poly(CG) 
sequence shown, the HOMO orbitals  are always
associated with the G sites for all the stretching modes, while
the LUMO orbitals are related to the C sites. 
However, as the DNA becomes more elongated, the orbitals overlap
even less and become localized for high stretching modes.
The elongation to the overstretched form is achieved by changing
the dihedral angle configuration of the DNA backbone, which leaves
the local part of the orbitals essentially intact. Note how the
orbitals rotate and spread out as the structure is being ovestretched,
following the rotation of bases. 

\begin{figure}
\includegraphics[width=3.5in]{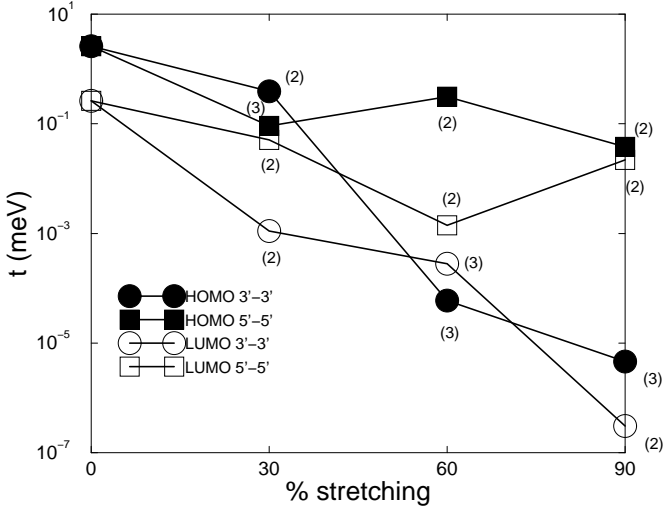}
\caption{The frontier state ``bottleneck'' hopping matrix 
elements as given by Eq.~(\ref{Eq:bottleneck}) for the different 
types (3'-3' or 5'-5') and amounts of stretching of poly(CG)-poly(CG) 
DNA. At each value of stretching, the dominant hopping process is 
indicated in parenthesis.} 
\label{Fig:hop}
\end{figure}

We now turn to a discussion of the results for the hopping matrix
elements of Eq.~(\ref{Eq:Hopping}). Our discussion here is relevant 
to what happens when the occupation of a frontier
state is changed from complete filling (for the HOMO) or complete
depletion (for the LUMO), that is, the physics of small amounts of
hole or electron doping. 
In Table \ref{Table:hoppings} we
give the values for $\varepsilon, t_1, t_2, t_3$ (see Fig.
\ref{Fig:schem}) for the two frontier states of the unstretched
poly(CG)-poly(CG) DNA form. The hopping matrix elements for the
HOMO state involve only the G sites; those for the LUMO state
involve only the C sites. As a consistency check, we have also calculated
matrix elements for farther neighbors and found those to be much
smaller in magnitude.
We have calculated the values of
$t_1,t_2,t_3$ by repeating the same procedure as above for the
stretched forms of the poly(CG)-poly(CG) DNA sequence. We note
that if $t_{2}=t_{3}=0$ electrons will not be able to migrate
along the DNA molecule even if $t_{1}$ is quite large, because at
least one of the other two hops is necessary for 
migration (see Fig. \ref{Fig:schem}). From this simple picture, it
is evident that the conductivity will be determined by which
matrix element dominates. Quantitatively, the ``bottleneck''
hopping matrix element is given by
\begin{equation} \label{Eq:bottleneck}
t=\max\left(\min(|t_{1}|,|t_{2}|),|t_{3}|\right).
\end{equation}
In Fig.~\ref{Fig:hop} we show the value of the ``bottleneck'' 
hopping matrix element calculated as a function of stretching. This
indicates that hopping conductivity will dramatically decrease by
several orders of magnitude upon stretching the molecule and that
the hopping will decrease more from stretching in the 3'-3' mode
than in the 5'-5' mode. This is due to the conformational changes
induced by the different stretching modes, described earlier.

\subsection{Localization length}

\begin{figure}
\includegraphics[width=3.2in]{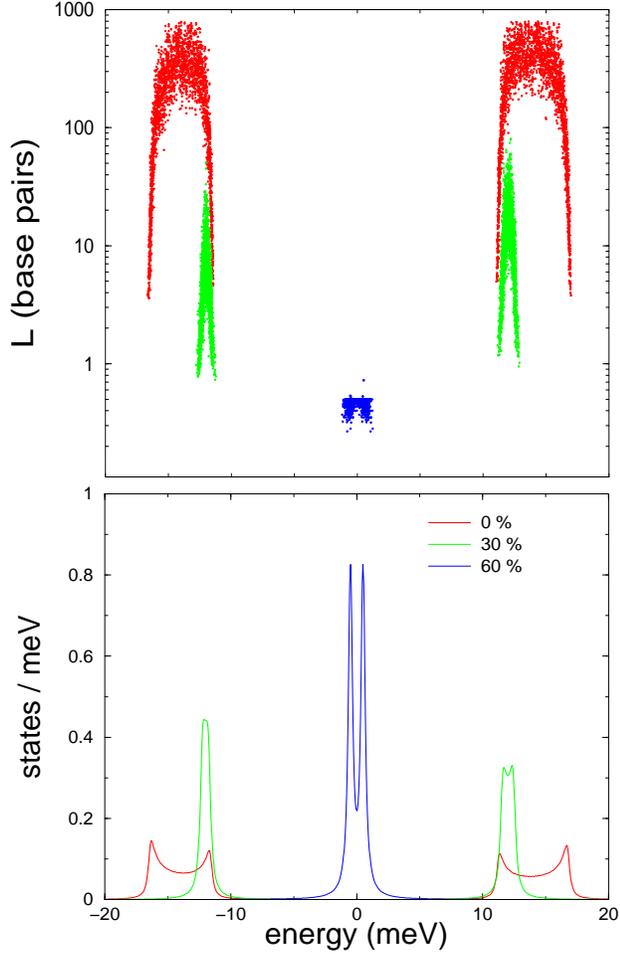}
\caption{(bottom) The density of electronic states for the HOMO
state stretched in the 3'-3' mode. For comparison, the on-site
energy parameter, $\varepsilon$, has been set to zero. (top) The
localization length $L_{i}$, defined in Eq.~(\ref{loc_length}), is 
computed for each eigenstate with disorder strength $\gamma=0.3$
meV.} \label{Fig:dos}
\end{figure}

The significant dropping of the hopping matrix elements upon
stretching as described in the previous section is indicative
of electron localization with a weak amount of disorder.  To
investigate this possibility in detail,
we focus on effects of stretching in
the 3'-3' mode. The evolution of
the density of HOMO states upon stretching is shown in
Fig.~\ref{Fig:dos}; similar behavior is observed for the LUMO
states. The dramatic narrowing of the DOS width (equivalent to
reduced 
dispersion in a band-structure picture) is strongly suggestive of
electron localization \cite{ANDERSON1958}, in this case induced by
stretching. 
This localization length is controlled by the hopping
elements $t$, since $\varepsilon$ is the same at each site.

For a more quantitative description,  
we show in Fig.~\ref{Fig:dos} the
localization length $L_{i}$  
for each eigenstate for a $1500$ base-pair DNA
strand under different amounts of stretching. 
The value of $L^{(i)}$ for each state is obtained from Eq.(\ref{loc_length}), 
with disorder
strength $\gamma=0.3$ meV, which determines the width of
the gaussian given in Eq.~(\ref{PU_disorder}).
This disorder strength is much smaller than the
band width of the unstretched DNA, but becomes comparable to the
band width as the molecule is stretched. The magnitude of such
variations in on-site energies is consistent with those produced
by the dipole potential terms, for instance, due to the presence
of a stray water molecule situated on the substrate roughly 15
\AA$ $ away from the DNA bases. 
We find that 
changing the value of $\gamma$ by an order of magnitude
(either smaller or larger) 
does not affect the qualitative picture presented here.
Note that the localization length is not a strict function of the
energy, as it depends on the disorder near where a given state
happens to be localized. As the molecule is stretched, the
localization length dramatically decreases until, for 60\%
stretching, the eigenstates are completely localized on single base pairs.

The charge localization length 
as a function of DNA stretching has been
recently studied in the
experiment of Heim {\it et al.}~\cite{Heim2004}. 
This study focuses on $\lambda$-DNA which
has an irregular sequence of base pairs, and can be compared 
to our theoretical results for poly(CG)-poly(CG) recalling
that the frontier states even for a random sequence are associated 
with those of the CG base-pairs. In the experiment,
ropes of $\lambda$-DNA on a substrate are overstretched by a
receding meniscus technique. The DNA ropes in this experimental
setup are slightly positively charged, corresponding to a
depletion of a few electrons per 1000 base pairs. 
We suggest that this situation is approximated by the 
structures of dry and neutral DNA that we considered above.
Electrons were
injected into the DNA and the resulting localization length was
measured by an electron force microscope. For the unstretched DNA,
the charge was found to delocalize across the entire molecule,
extending over a length of several microns. On the other hand, the
charge injected into the overstretched DNA is localized, extending
over a few hundred nanometers only.
This is qualitatively consistent with the picture that emerges
from our theoretical analysis, and is even in reasonable quantitative
agreement: the degree of localization in experiment, measured by
the ratio of length scales going from unstretched to stretched
DNA structures, is approximately two orders of magnitude, 
while the same quantity in our calculations, going from unstretched
to 60\% stretched DNA is $\sim 10^3$.

\section{Summary}

We have described and implemented a multiscale
method to derive effective hamiltonian models that are able to
capture the dynamics of conduction and valence electrons in
stretched DNA, starting from {\it ab initio}, all-atom quantum
mechanical calculations.
The {\it ab initio} simulations revealed that the frontier states
in the base pairs are related to only one component of the pair.
The purines were found to be associated with the HOMO states
while the pyrimidines with the LUMO states. In the AT-CG combination
the frontier states are identified with those of the CG pair.
For all combinations of bases and base pairs studied here, the 
nature of these 
states was not affected by separation of the bases 
or base pairs along different directions or rotation along the helical axis.

Turning to the next length scale and the semi-empirical calculations,
we have calculated the ``bottleneck'' matrix elements
for electron hopping along the DNA molecule, as a function of stretching.
These show a significant decrease with elongation of DNA, 
which is stronger for stretching 
in the 3'-3' mode than in the 5'-5' mode. We were able to show
{\em quantitatively} that stretching of DNA dramatically narrows
the DOS width of frontier states. A small amount of
 disorder produced by environmental factors will naturally lead to
 localization of the electrons along the DNA.  Our estimate for the
 degree of localization, based on a reasonable (and quite small)
 amount of disorder in the on-site energies for the electron states, is
 in very good agreement with recent experimental observations. This
provides direct validation for the consistency and completeness of
the multiscale method presented here.

{\bf Acknowledgements:} The authors are grateful to Richard Lavery
for providing the overstretched structures. 
MF acknowledges support by Harvard's Nanoscale Science and 
Engineering Center, funded by the National Science Foundation, 
Award Number PHY-0117795.

\newpage

\end{document}